\documentclass[mathleft]{an}

\usepackage{graphicx}
\usepackage{amsmath,amsfonts,amssymb}
\usepackage{times}

\def\rot{\mathop{\rm rot}\nolimits}
\def\div{\mathop{\rm div}\nolimits} 
\def\gsim{\lower.4ex\hbox{$\;\buildrel >\over{\scriptstyle\sim}\;$}} 
\def\lsim{\lower.4ex\hbox{$\;\buildrel <\over{\scriptstyle\sim}\;$}} 

\def\[{\begin{eqnarray}}
\def\]{\end{eqnarray}}

\def\q{\qquad}

\newcommand{\nablasq}{\nabla^2}

\newcommand{\Rey}{\mathrm{Re}}

\newcommand{\Rm}{\mathrm{Rm}}
\newcommand{\Omin}{\Omega_{\mathrm{in}}}
\newcommand{\Omout}{\Omega_{\mathrm{out}}}
\newcommand{\Rin}{R_\mathrm{in}}
\newcommand{\Rout}{R_\mathrm{out}}

\newcommand{\muh}{\hat{\mu}}
\newcommand{\etah}{\hat{\eta}}

\newcommand{\Pm}{\mathrm{Pm}}
\newcommand{\Ha}{\mathrm{Ha}}

\newcommand{\ord}[1]{$O(#1)$}

\newcommand{\mperm}{\mu_0}
\newcommand{\mdiff}{\eta}

\newcommand{\cmnt}[1]{}
\newcommand{\comm}[1]{}
\newcommand{\ignore}[1]{}

\begin{document}

\setlength{\mathindent}{0cm}
\sloppy


 \Pagespan{844}{}
 \Yearpublication{2006}%
 \Yearsubmission{2006}%
 \Month{11}%
 \Volume{327}%
 \Issue{9}%
 \DOI{10.1002/asna.200610662}%

 \title{Nonlinear simulations of magnetic Taylor-Couette flow with current-free 
 helical magnetic fields }

 \author{J. Szklarski\thanks{Corresponding author:
   jszklarski@aip.de} \and G. R\"udiger}

 \institute{
 Astrophysikalisches Institut Potsdam, An der Sternwarte 16, 
 D-14482 Potsdam, Germany} 
 \authorrunning{J. Szklarski \and G. R\"udiger}
 \titlerunning{Numerical simulations of MHD TC flow with helical magnetic fields} 
 \received{July 2, 2006}
 \accepted{later}
 \publonline{more later}

 \abstract{The  magnetorotational instability (MRI) in cylindrical 
 Taylor-Couette flow with external helical magnetic field is simulated for infinite and
 finite aspect ratios.  We solve the MHD equations in their small Prandtl number
 limit and confirm with time-dependent nonlinear simulations that the
 additional toroidal component of the magnetic field reduces the critical
 Reynolds number from \ord{10^6} (axial field only) to \ord{10^3} 
 for  liquid metals with their small magnetic Prandtl number. Computing
 the saturated state  we obtain velocity amplitudes
 which help designing  proper experimental setups. Experiments with
 liquid gallium require axial field $\sim 50$ Gauss and axial current
 $\sim 4$ kA for the toroidal field. It is sufficient that the vertical velocity $u_z$ of the flow can be measured with a precision of 0.1mm/s.\\
 We also show that the endplates enclosing the cylinders do not destroy
 the traveling wave instability which can be observed as
 presented in earlier studies.  For TC containers without and with endplates the angular momentum transport of the MRI instability is shown as to be outwards.}


 \maketitle

 \section{Introduction}

 The magnetorotational instability (MRI) has been formulated long time
 ago (Velikhov 1959) but recently it become of particular
 interest for astrophysics due to recognition of its importance as
 the source of turbulence in accretion disks with Keplerian rotation
(Balbus \& Hawley 1991). One of the most convenient models
 for MRI is magnetohydrodynamical cylindrical Taylor-Couette flow, i.e.
 motion of liquid metal between two concentric differentially rotating
 cylinders under imposed external magnetic field along the axis (R\"udiger \& Zhang 2001; Ji, Goodman \& Kageyama 2001; Willis \& Barenghi 2002). 
Under the influence of a purely
 axial field MRI operates for magnetic Reynolds number $\Rm=\Omin
 \Rin(\Rout-\Rin)/\mdiff$ of order \ord{10} and has never been observed in an
 experiment due to very small Prandtl number for laboratory liquids (see
 Table~\ref{tab-physprop}) which results in need of 
 high critical Reynolds numbers of order $10^6$  .
 Recently it has been shown by Hollerbach \& R\"udiger (2005) and R\"udiger et al. (2005) 
 that a current-free external toroidal magnetic field in addition to the  usual axial field can
 reduce the critical Reynolds number  to \ord{10^3} and therefore
 it makes it much easier to design a MRI experiment due to slower
 rotation.

 The aim of the presented paper is to compute the amplitudes of
 velocities and the frequencies in a nonlinear saturated state under the influence  of the
 endplates covering the cylinders.
  
 \begin{figure}
    \center \includegraphics[width=6.0cm]{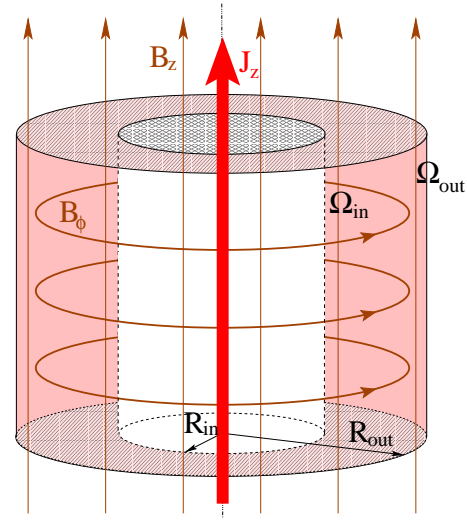}
    \caption{The geometry of the problem -- two concentric cylinders with radii
	     $\Rin=4$ cm, $\Rout=8$ cm rotating with $\Omin$,
	     $\Omout$. $B_z$ and $B_\phi$ are the external magnetic
	     fields. $B_\phi$ is due to an axial current inside the inner
	     cylinder. In the PROMISE experiment cylinders have finite
	     length of $H=40$ cm ($\Gamma=10$) and are covered with
	     endplates. The liquid is a mixture of gallium, indium
	     and tin (Stefani et al. 2006).}
    \label{fig-geom}
 \end{figure}

 The fluid confined between the cylinders is assumed to be incompressible and
 dissipative with the kinematic viscosity $\nu$ and the magnetic diffusivity
 $\mdiff$. Derived from  the conservation of angular momentum the rotation law
 $\Omega(R)$ in the fluid is
 \[
    \Omega(R)=a+\frac{b}{R^2}
 \]
 with
 \[
  a=\frac{\muh-\etah^2}{1-\etah^2} \Omin, \q 
  b=\frac{1-\muh}{1-\etah^2}\Rin^2 \Omin,
 \]
 where $\muh=\Omout/\Omin$. After the Rayleigh stability criterion, i.e. 
 $\partial_R(R^2\Omega)^2 > 0$, the flow is hydrodynamically stable for
 $\muh>\etah^2$, i.e.  $\muh>0.25$ for our geometry.
 In order to isolate MRI we are mainly interested in hydrodynamically
 stable regimes so that  $\muh>\etah^2$ should be  fulfilled.

\section{Equations and numerical model}

 The dimensionless incompressible MHD equations are
 \begin{eqnarray}
 \lefteqn{       \partial_t \vec{u} + (\vec{u} \cdot \nabla)\vec{u} =
               -\nabla p + \nablasq \vec{u} + \frac{\Ha^2}{\Pm} 
               (\rot \vec{B}) \times \vec{B},} \nonumber\\
\lefteqn{       \partial_t \vec{B} =
                \frac{1}{\Pm} \nablasq \vec{B} + 
	        \rot (\vec{u} \times \vec{B}),} 
\end{eqnarray}
with       $\div{\vec{u}} =  \div{\vec{B}} = 0$ and
  with the Hartmann number
 \[
    \Ha = B_0 \sqrt{\frac{\Rin (\Rout-\Rin)}{\mperm \rho \nu \mdiff}}.
     \]
The magnetic Prandtl number is $\Pm = \nu/\mdiff$. 
  $D=\Rout - \Rin$ is used as the unit of length,  $\nu/D$ as the 
 unit of velocity and $B_0$ as the unit of magnetic fields. The Reynolds number
 $\Rey$ is defined as $\Rey=\Omin \Rin D/ \nu$.
 
 \begin{table}
    \tabcolsep5pt
    \caption{Physical properties of liquid metals which are suitable for 
             the MRI experiment. A star denotes the PROMISE mixture.}
    \begin{tabular}{lcccc}
       \hline
        & $\rho$ [g/cm$^3$] & $\nu$ [cm$^2$/s] & $\mdiff$ [cm$^2$/s] & $\Pm$ \\[0.5ex]
       \hline\\[-8pt]
        sodium & 0.9 & $7.1\cdot 10^{-3}$ & $0.8\cdot 10^3$ &$0.9\cdot 10^{-5}$ \\[0.5ex]
        gallium & 6.0 & $3.2\cdot 10^{-3}$ & $2.1\cdot 10^3$ &$1.5 \cdot 10^{-6}$ \\[0.5ex]
	gallium$^{*}$ & 6.4 & $3.4\cdot 10^{-3}$ & $2.4\cdot 10^3$ &$1.4 \cdot 10^{-6}$ \\[0.5ex]
       \hline
    \end{tabular}
    \label{tab-physprop}
 \end{table}

 For laboratory liquids like gallium, the magnetic Prandtl number is very small 
 (see Table~\ref{tab-physprop}) so the magnetic diffusion time is 
 much shorter than other time scales and fluctuations $\vec{b}$ of the
 field $\vec{B_0}+\vec{b}$ are also much smaller than the applied field $\vec{B}_0$. 
 As a result the $\vec{b}$ adjusts instantaneously to the velocity $\vec{u}$ 
 and the quasi-static approximation can be used 
(Roberts 1967; Zikanov \& Thess 1998;  Youd \& Barenghi 2006).
 
 With  $\vec{B} \rightarrow \vec{B_0} + \Pm \vec{b}$ and
 $\Pm \rightarrow 0$ one finds the equation system
 \begin{eqnarray}
   \label{eqn-quasimhd}
\lefteqn{ \partial_t \vec{u} + (\vec{u} \cdot \nabla)\vec{u} =
       -\nabla p + \nablasq \vec{u} + 
       \Ha^2 (\nabla \times \vec{b}) \times \vec{B_0},} \nonumber\\
\lefteqn{ \nablasq \vec{b} = - \rot (\vec{u} \times \vec{B_0})} 
 \end{eqnarray}
and
$
 \div{\vec{u}}=  \div{\vec{b}}= 0.  
 $
The external field is  
 \[
   \vec{B_0} = (0, B_\phi, B_0) = B_0\left(0, \frac{\beta \Rin}{R}, 1\right),
 \] 
where the  parameter $\beta$ denotes the ratio of toroidal field at the inner cylinder
 to the constant axial field $B_0$. Except along the axis the magnetic field is current-free.
 
 We simulate the axisymmetric 2D flow in cylindrical coordinates 
 ($R, \phi, z$) with the numerical code of A. Youd (see Youd \& Barenghi 2006 
 for details). The code has been modified in order
 to handle periodic boundary conditions in a way suitable for our needs, 
 the toroidal  field was added and different boundary conditions on the 
 endplates were applied. 
 
 Equations (\ref{eqn-quasimhd}) are solved with a finite-difference 
 method in the $R$-$z$ plane in a streamfunction-vorticity formulation, i.e.
 \begin{eqnarray}
\lefteqn{ \partial_t u_\phi = (\nablasq -R^{-2})u_\phi + 
       [\vec{u} \times (\rot{\vec{u}})]_\phi + 
        \Ha^2 \partial_z B_{\phi}} \nonumber \\
\lefteqn{ \partial_t \omega_\phi = (\nablasq -R^{-2})\omega_\phi -} \nonumber\\
    && \qquad\qquad\quad  [\rot{(\vec{\omega}\times\vec{u})}]_\phi + 
       \Ha^2[\rot{(\vec{j}\times \vec{B_0})}]_\phi,
 \end{eqnarray}
  with  `Poisson equations' for streamfunction, magnetic field and the current
\begin{eqnarray} 
\label{poissons}
    \lefteqn{ \left(-R^{-1}(\partial_{zz} +\partial_{rr}) + 
       R^{-2}\partial_r \right) \psi =  \omega_\phi} \nonumber \\
    \lefteqn{ (\nablasq \vec{B})_\phi = 
       -\partial_z (u_\phi - u_z \beta B_0 \frac{\Rin}{R}) +
     \partial_R u_R \beta B_0 \frac{\Rin}{R}} \nonumber\\
    \lefteqn{ (\nablasq \vec{j})_\phi = 
       - \partial_{zz} (u_R) + \partial_R  \frac{\partial_R(R u_R )}{R},}
 \end{eqnarray}
 where the streamfunction $\psi$ is defined by $u_R = -(1/R)\partial_z \psi$, 
 $u_z = (1/R)\partial_R \psi$ and vorticity $\vec{\omega}=\rot(\vec{u})$.
 
 No-slip boundary conditions for the velocity on the walls and the
 endplates are used, so that $u_r=0$ at $R=\Rin, \Rout$ and $u_z=0$
 at $z=0, H$.  Therefore  $\psi=0$ at $R=\Rin, \Rout$ and $z=0, H$. Using
 (\ref{poissons})$_1$ 
 we obtain $\omega_\phi=-(1/R)\partial_{rr}\psi$ at $R=\Rin,
 \Rout$ and $\omega_\phi=-(1/R)\partial_{zz}\psi$ at $z=0, H$.  Boundary
 conditions for $u_\phi$ at the walls are determined by $\Rey$ and $\muh$
 and at the endplates by their rotation properties -- fixed or rotating
 as rigid body with $\Omout$ or $\Omin$.

 The magnetic boundary conditions depend on the electrical properties of 
 the walls. For perfectly conducting cylinders we assume infinite 
 conductivity so that with $\vec{J}=
 \sigma(\vec{E}+\vec{u} \times \vec{b})$ and $E_\phi=E_z=0$ for the walls 
 we get
 \[
    R^{-1}b_\phi + \partial_R b_\phi=
     0, \quad j_\phi=0 \quad \mathrm{at} \quad R=\Rin,\Rout.
 \]  
 In the present paper only insulating endplates are 
 considered and the BCs for magnetic field are obtained by 
 applying the so called pseudo-vacuum approximation for which 
 $b_r=b_\phi= \partial_z b_z=0$. Then
 \[
    b_\phi=  \partial_z j_\phi=0
 \]
at $z=0, H$.
 We use grid resolution $N_R \times N_z=80 \times 800$ (and $40 \times
 400$ for comparison, the difference in obtained values is always less
 than couple of percent). The time-step is constant with d$t=10^{-5}$ or d$t=10^{-4}$.

\section{Infinite container}

 First we have performed simulations for periodic vertical boundary
 conditions in order to compare the critical Reynolds numbers
 with those presented in R\"udiger et al. (2005) as well as to
 compute for the first time amplitudes of velocities in the nonlinear
 saturated state.  For infinite cylinders and $\muh=0.27$ the flow is
 always hydrodynamically stable, with external axial magnetic field
 it looses the stability for large Reynolds number of order \ord{10^6}
 but the additional toroidal current-free field can reduce this number to 
 \ord{10^3} -- for $\beta=3, 4$ the critical $\Rey_{\rm crit}$ is $1160$ and
 $842$, resp.

 In the simulations the length of the periodic cylinders was chosen
 to be three times the wavelength obtained with linear analysis. The
 agreement between the previous and presented results is rather good.

 \subsection{Frequencies}

 \begin{figure}
    \center
    \includegraphics[width=8.5cm]{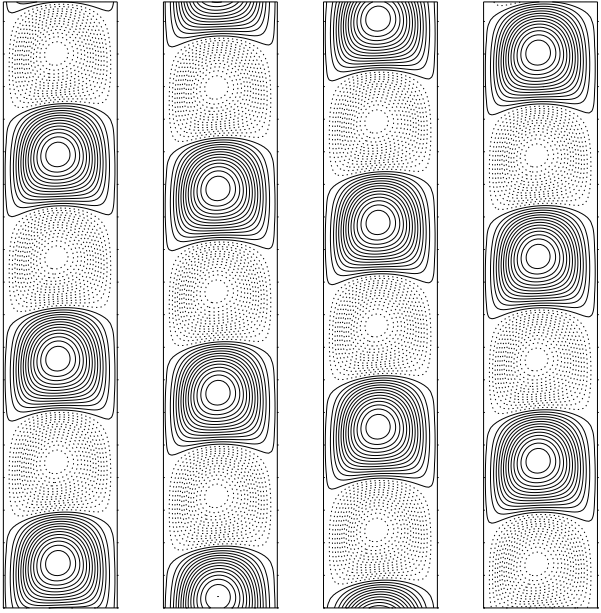}
    \caption{Snapshots of contour lines of the streamfunction: dashed lines 
         represent counterclockwise rotation,
	 solid lines clockwise rotation.  
	 $\Ha=9.5, \Rey=900, \muh=0.27, \beta=4$. 
	 The graph is not at scale.}
    \label{fig-isnap}
 \end{figure}

 \begin{figure}
    \center
    \includegraphics[width=8.8cm]{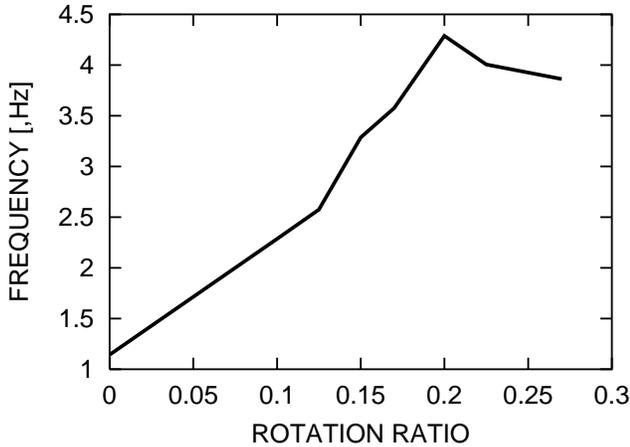}
    \caption{Frequency of the traveling wave as a function of
	     rotation ratio $\hat\mu$ for infinite cylinders, $\Ha=9.5, \Rey=900,
	      \beta=4$.  Here and below the units are scaled
	     according to physical properties of gallium.}
    \label{fig-ifreq}
 \end{figure}
 
 Figure~\ref{fig-isnap} shows snapshots of the streamfunction for the saturated
 state for Reynolds number just above the critical value, the three
 pairs of drifting vortices can be seen. This drift and its velocity
 agrees with the previous results, it is slow, for 
 gallium the period of the traveling wave is about 230 s for $\beta=4$ which
 corresponds to $u_{\mathrm{drift}}=0.5$ mm/s (the period of one rotation
 of the inner cylinder is about 35 s for $\Rey=900$).

 From the experimental point of view it is also convenient to present
 traveling wave frequencies as function of $\muh$ (Fig.~\ref{fig-ifreq}).  All the frequencies were obtained by taking
 the spectrum of the velocity component $u_z(t)$ in the middle of the gap and
 then choosing the dominant frequency. Except for very small $\muh$ the resulting frequencies are always smaller 
 than the frequencies of both the cylinders. Beyond the Rayleigh limit (i.e. for $\muh > 0.25$) the 
 characteristic frequency is about 10\%  of the frequency of the inner cylinder. That means that beyond the 
 timescale of rotation another timescale  exists in magnetic  astrophysical systems which exceeds the rotation period 
 by (say) a factor of ten. Due to the small electric conductivity, however, our TC-system only possesses a 
 very short diffusion time of order of  10$^{-2}$s.


\subsection{Velocity amplitudes and torque}

 \begin{figure}
    \center
    \includegraphics[width=8.8cm]{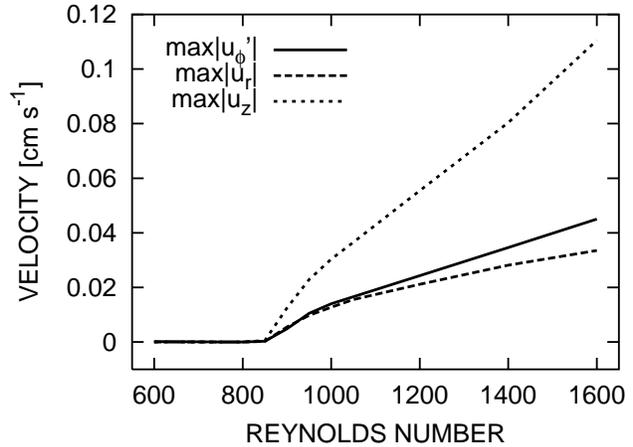}
    \caption{Amplitudes of velocity as function of Reynolds number, 
    $u_\phi'$ is the deviation from the standard Couette solution. The amplitudes
    increase from the critical point $\Rey_c=842$ and continue to increase with $\Rey$. 
    Infinite cylinders, $\Ha=9.5, \muh=0.27, \beta=4$.} 
    \label{fig-amplre}
 \end{figure}

 \begin{figure}
    \center
    \includegraphics[width=8.8cm]{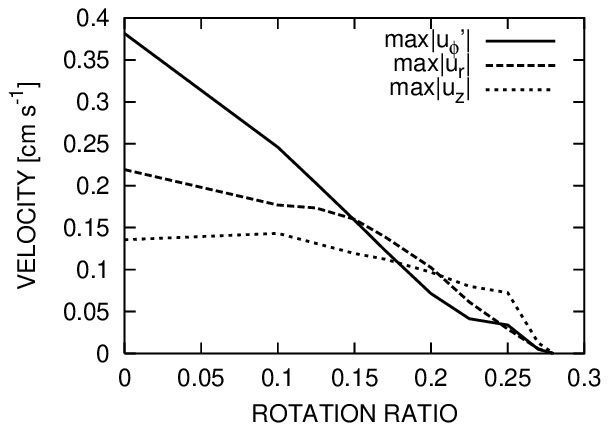}
    \caption{Average maximum amplitudes of the velocity versus $\muh$ for 
             infinite cylinders, $\Rey=900, \beta=4, \Ha=9.5$. 
	     \comm{$u_\phi'$ is deviation from stable Couette solution.}
	     } 
    \label{fig-iamplu}
 \end{figure}

 The saturated nonlinear solution provides the values of velocity amplitudes
 which have not been known before. The velocity component which
 is easiest to measure in the laboratory is $u_z$.

 Figure~\ref{fig-amplre} shows the dependence of different components
 of velocity as function of Reynolds number $\Rey$. The velocities are
 zero up to critical point $\Rey_{\rm crit}$ at which they start to grow. We
 notice that the values of $\vec{u}$ can be further increased by increasing
the rotation rates of the container and therefore it is easier to measure $u_z$
 for higher $\Rey$.

 In Fig.~\ref{fig-iamplu} the maximum amplitudes of velocity
 for $\Rey=900$ and $\beta=4$ are presented. For any stable solution all the
 amplitudes should be zero just beyond the Rayleigh line, i.e. on the
 right of $\muh=0.25$. We see that with additional toroidal field this
 is not the case -- the velocities for $\muh=0.27$ are of order 0.1 mm/s
 and drop here to zero for $\muh=0.28$ for these parameters.

 \begin{figure}
    \center
    \includegraphics[width=8.8cm]{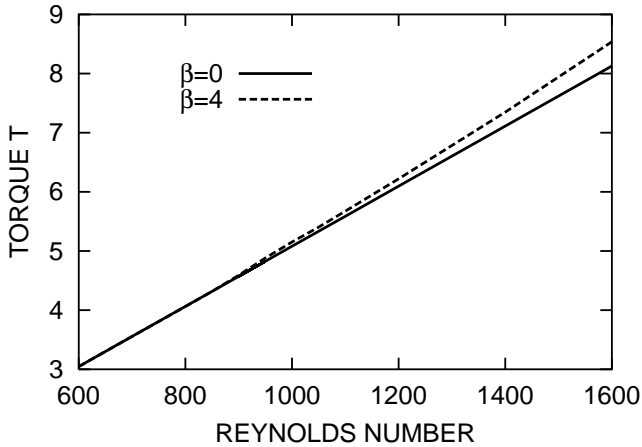}
    \caption{Torque at the walls for two
	     different $\beta$. Note that from the critical 
	     $\Rey_{\rm crit}=842$ for  $\beta=4$ case the torque is
	     larger than for the stable $\beta=0$ case, $\Ha=9.5$.}
    \label{fig-torqueinf}
 \end{figure}

 For MRI one expects the angular momentum to
 be transported outwards and this should result in an increase of the torque.
 We are mainly interested in torques at the walls at which the magnetic
 and advective torque vanish so the total torque is equal to the
 viscous one, which is defined as
 \[
  T(R)=-\int_0^H 
     R^3 \rho \nu \partial_R \left( \frac{u_\phi(R)}{R}\right)  \ \mathrm{d} z.
 \]
 Although the change of the torque is rather small and would be difficult to
 measure in the laboratory, Fig.~\ref{fig-torqueinf} shows that 
 increase of the torque can be observed in our computations above the critical Reynolds number. Indeed the angular momentum is transported outwards.

\section{Finite container}

 In any real experiment finite cylinders enclosed by
 endplates are used.
 It has been pointed out that the endplates can have significant impact on
 the flow especially for short cylinders and for fast rotation
(e.g. Hollerbach \& Fournier 2004; Kageyama et al. 2004)
 also some solutions for reduction of the endplates effect have 
 been proposed  
 (Burin et al. 2006).
We are mainly interested in knowing
 how the endplates and their mechanical and magnetic properties 
 influence the velocity amplitudes and frequencies for slow rotating 
 finite cylinders with aspect ratio of (say) $\Gamma=10$ and a current-free helical external field geometry. 
 
\subsection{Ekman layer}


 The existence of the viscous endplates results in developing so-called
 Ekman layers in which the velocity differs from any toroidal rotation
 as centrifugal force is weakened close to the vertical boundaries. 
 This effect drives a global meridional circulation and
 two Ekman vortices appear.
 The velocity in this region depends on the Reynolds number and
 the velocity of the endplates themselves. As known for nonrotating
 endplates (or rotating as a solid body with $\Omout$) there is a radial
 inflow close to the boundaries and for solid-body rotation with $\Omin$
 there is a radial outflow.

 In the Rayleigh-stable regime the flow is hydrodynamically stable for
 both the infinite and finite containers, whereas for the latter two
 Ekman vortices are always present. We confirmed it with the numerical
 method that it is used here.

 A simple estimate of the thickness of the Ekman layer (for endplates
 rigidly rotating with $\Omout$, see e.g. 
Ji et al. 2004)
 gives 
\[
\delta \approx \sqrt{\frac{\nu}{\bar{\Omega}}} \approx 0.13\ {\rm cm} 
\label{delE}
\]
 for the $\beta=4$ critical case where
 $\bar{\Omega}=\sqrt{\Omin \Omout}$. This is a larger value in comparison
 to other experiments dealing with faster rotation ($\Rey \approx 10^6$)
 in which case the possibility of developing hydrodynamical turbulence
 in the Ekman layer arises.

 The typical velocity in the Ekman flow can be approximated by
 $\bar{u_\delta}=\sqrt{\mu \bar{\Omega}}$ which for $\Rey=900$ with 0.02
 cm/s  is of the same order for velocities for the unstable case presented for infinite
 cylinders (see Fig.~\ref{fig-iamplu}). For $\Rey=900$ the maximum value
 of $u_z$ (in the corners) is above 0.1 cm s$^{-1}$ which is ten times
 more than the values computed for infinite cylinders. 
 Moreover, for the same parameters the Ekman circulation time $H/2\sqrt{\nu \bar{\Omega}}$ is of order 
 $10^3$s which is comparable to typical MRI growth time.
 We must
 conclude that endplates can significantly alter the flow.

 In the computations it turns out that in the presence of the external
 axial field, the Ekman flow induces significant $z$-gradients in $b_\phi$
 which can result in a magnetic instability even for small Reynolds number. The critical
 numbers depend on magnetic properties of the endplates, their rotation
 and Hartmann number\footnote{e.g. $\Rey_{\rm crit}=140$ for $\Ha=3.5$, perfectly conducting, fixed 
 endplates}. Since at this moment our aim is to investigate how the traveling-wave  effects
 of the MRI with helical (current-free) field structure can be observed experimentally, this
 instability, although highly interesting, is beyond the scope of the present paper.

 \begin{figure}
    \center
    \includegraphics[width=8.8cm]{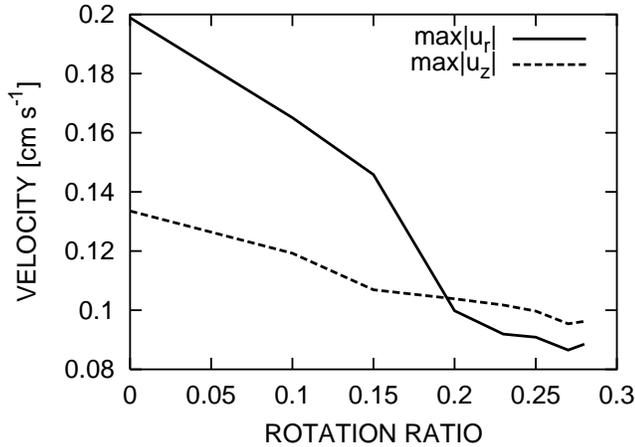}
    \caption{The same as in 
             Fig.~\ref{fig-iamplu} but for finite cylinders. $\Rey=900, \beta=4, \Ha=9.5$,  
             insulating endplates (pseudo-vacuum boundary 
	     conditions) both rotating with the outer cylinder.} 
    \label{fig-famplu}
 \end{figure}
 
 \begin{figure}
    \center
    \includegraphics[width=8.8cm]{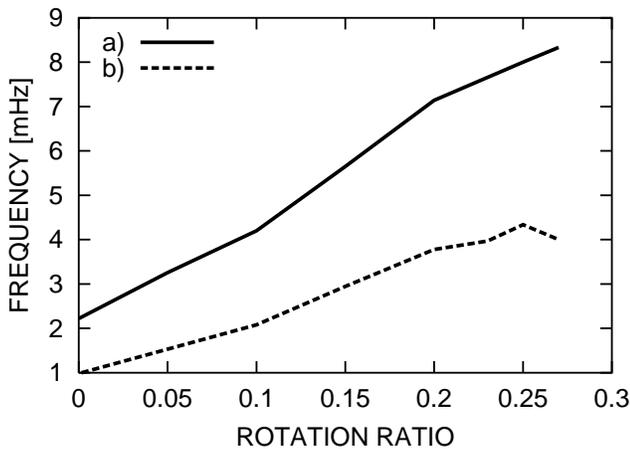}
    \caption{Frequencies as function of rotation ratio $\muh$ for
    finite cylinders. a) $\beta=6, \Rey=1500, \Ha=9.52$, insulating endplates, 
                bottom: fixed, upper: rotating with $\Omout$. 
	     b) $\beta=4, \Rey=900, \Ha=9.5$, insulating endplates 
	        rotating with $\Omout$.} 
    \label{fig-fcmp}
 \end{figure}

 \begin{figure}
    \center
    \includegraphics[width=8.8cm]{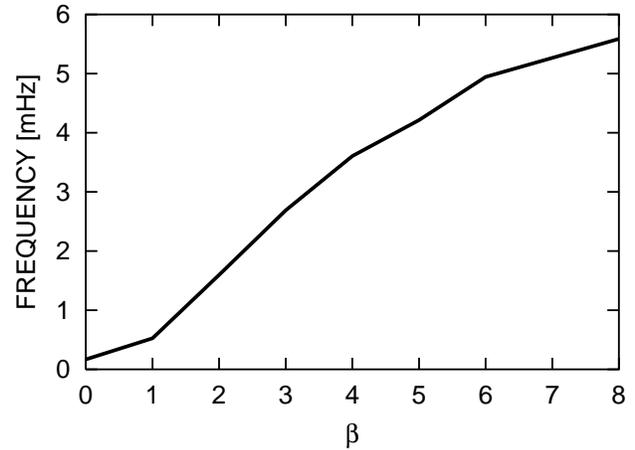}
    \caption{Frequencies of the traveling waves in the middle of the 
             gap for finite cylinders as function of 
	     $\beta$. $\Rey=900, \muh=0.27, \Ha=9.5$, 
	     insulating endplates rotating with $\Omout$.}  
    \label{fig-ffreqbeta}
 \end{figure}

\subsection{Amplitudes and frequencies}

 The velocity amplitudes for finite cylinders just above the critical Reynolds
 number are shown in Fig.~\ref{fig-famplu}.  We notice that due to the Ekman circulation those
 velocities do {\em not} drop to zero even for $\muh=0.28$, which was true
 for infinite cylinders.
Again the external toroidal
 field maintains traveling waves which can easily be observed in the
 laboratory.
  Analog to Fig.~\ref{fig-ifreq}, Fig.~\ref{fig-fcmp}
 shows the frequencies of the velocity field in the center of the
 gap. Unlike for the infinite container, here the position of measurement does
 make difference since in the vicinity of the endplates the traveling
 wave is suppressed by the Ekman layer. 
 From Fig.~\ref{fig-ffreqbeta} we
 can clearly see that non-zero $\beta$ introduces vertical movement with
 very similar frequencies as for the infinite case (compare with $\muh=0.27$ in
 Fig.~\ref{fig-ifreq}).  

 \begin{figure}
    \center
    \includegraphics[width=8.5cm]{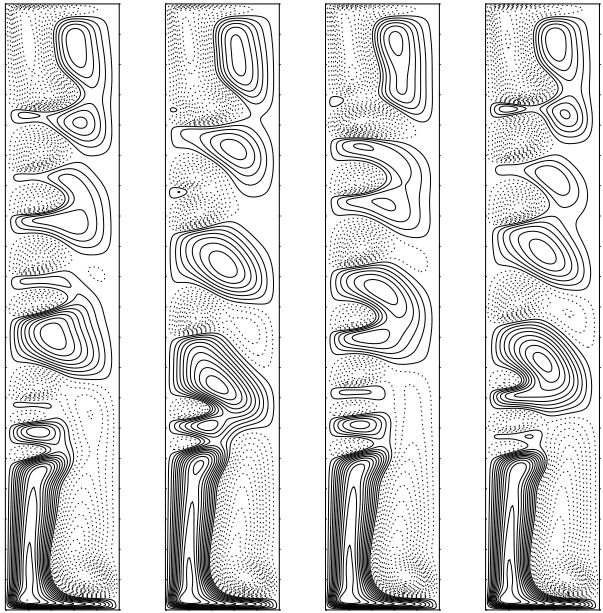}
    \caption{Snapshots of contour lines of the streamfunction analog to 
             Fig.~\ref{fig-isnap} but for finite cylinders. $\Rey=1500, \beta=6, \Ha=9.5$, 
	     insulating endplates, bottom: fixed, upper: rotating with $\Omout$.}    
    \label{fig-fsnap}
 \end{figure}

 Snapshots of streamfunction are shown in Fig.~\ref{fig-fsnap}. We
 see that the flow is complicated and deformed when compared with infinite-container
 solutions. The $z$-asymmetry close to the endplates is due to different
 boundary conditions: the bottom endplate is at rest, the upper moves
 with the outer cylinder\footnote{This is the case for the current experimental
 setup of PROMISE}. We conclude that even for nonsymmetric rotation of the endplates
 the traveling wave can still be observed.

 \begin{figure}
    \center
    \includegraphics[width=8.8cm]{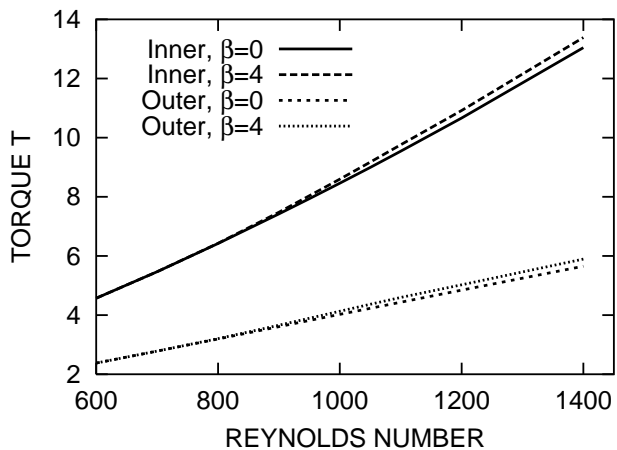}
    \caption{Torque at the walls (inner and outer) for {\em finite} cylinders
             for different values of $\beta$. $\Rey=900, \Ha=9.5$.}
    \label{fig-torquefin}
 \end{figure}

  
 Similarly like for the infinite case one can also observe the increase
 of torque in comparison with $\beta=0$ case (Fig.~\ref{fig-torquefin}).

\section{Conclusions}

We have shown that, as predicted, additional toroidal
 fields reduce the critical Reynolds number for the MHD Taylor-Couette
 flows for laboratory liquids with $\Pm \approx 10^{-5}-10^{-6}$. The
 nonlinear, axisymmetric simulations in the quasi-static approximation
 yield very similar characteristic values compared to those obtained with linear theory for
 infinite cylinders.
 
 The velocity amplitudes which were not known until now, are small
 but still it is not very difficult to measure them experimentally. Similarly
 the velocity of the drift can easily be detected by looking at changes
 in the velocity field. We shall also mention that increasing Reynolds
 numbers result in linear (at least to some point) increase of velocity
 amplitudes, therefore it is easier to detect them for higher $\Rey$,
 however for too high rotation rates the assumption of axisymmetry may not
 be valid.

 For finite aspect ratio calculations, the velocities induced by
 Ekman pumping are of similar order as those resulting from the
 instability. Therefore one should not neglect the influence of the
 endplates in such experiments. Nevertheless the influence of the toroidal
 field and the traveling wave can easily be detected in all cases.


\acknowledgements
J. Szklarski would like to thank A. Youd for providing the orginal version
of the code and for the discussions.


\end{document}